\documentclass{elsart}

\usepackage{natbib}

\usepackage{graphicx}
\usepackage{amssymb}

\begin{document}

\begin{frontmatter}

% Title, authors and addresses

% use the thanksref command within \title, \author or \address for footnotes;
% use the corauthref command within \author for corresponding author footnotes;
% use the ead command for the email address,
% and the form \ead[url] for the home page:
% \title{Title\thanksref{label1}}
% \thanks[label1]{}
% \author{Name\corauthref{cor1}\thanksref{label2}}
% \ead{email address}
% \ead[url]{home page}
% \thanks[label2]{}
% \corauth[cor1]{}
% \address{Address\thanksref{label3}}
% \thanks[label3]{}

\title{Hosts of Type II Quasars: an HST Study}

% use optional labels to link authors explicitly to addresses:
% \author[label1,label2]{}
% \address[label1]{}
% \address[label2]{}

\author{N.L.Zakamska, M.A.Strauss, J.H.Krolik, S.E.Ridgway}
\author{G.D.Schmidt, P.S.Smith, L.Hao, T.M.Heckman, D.P.Schneider}

%\address{}

\begin{abstract}
Type II quasars are luminous Active Galactic Nuclei whose centers 
are obscured by large amounts of gas and dust. In this contribution 
we present 3-band HST images of nine type II quasars with redshifts 
$0.25<z<0.4$ selected from the Sloan Digital Sky Survey based on 
their emission line properties. The intrinsic luminosities of these 
quasars are thought to be in the range $-24>M_B>-26$, but optical 
obscuration implies that host galaxies can be studied unencumbered 
by bright nuclei. Each object has been imaged in three filters 
(`red', `green' and `blue') placed between the strong emission 
lines. The spectacular, high quality images reveal a wealth of 
details about the structure of the host galaxies and their 
environments. Most galaxies in the sample are ellipticals, but 
strong deviations from de Vaucouleurs profiles are found, especially 
in the blue band. We argue that most of these deviations are due to 
the light from the nucleus scattered off interstellar material in 
the host galaxy. This scattered component can make a significant 
contribution to the broad-band flux and complicates the analysis 
of the colors of the stellar populations in the host galaxy. This 
extended component can be difficult to notice in unobscured luminous 
quasars and may bias the results of host galaxy studies.
\end{abstract}


\end{frontmatter}

One of the key long-term questions in quasar studies is to identify just which properties of the galactic host (overall morphology? presence of bars? close neighbors?  star-formation history? gas content?) are associated with activity in its nucleus. However, observations of quasar hosts are made difficult by the bright source in the center of the host galaxy -- the quasar itself; in some cases the light from the quasar completely overwhelms the host galaxy, preventing its detection altogether. Therefore, a significant advantage can be gained by studying the hosts of quasars in which the optical and UV light from the nuclear source are obscured by large amounts of gas and dust near the nucleus -- so called type II quasars \citep{urry95}. Such objects have only recently been discovered in large numbers in X-ray, mid-IR and optical surveys \citep[e.g.,][]{zaka03, lacy04, szok04}. 

In the optical, the strong blue continuum and broad emission lines typical of unobscured (type I) quasars are not observed in type II quasars, so that narrow emission lines that originate above and below the plane of obscuration are the only signature of the presence of a hidden Active Galactic Nucleus (AGN). We therefore searched the spectroscopic database of the Sloan Digital Sky Survey \citep[SDSS;][]{york00, abaz05} for objects with narrow emission lines with high-ionization line ratios and selected a few hundred type II quasar candidates in the redshift range $0.3<z<0.8$ \citep{zaka03}. We have been conducting sensitive follow-up observations of the objects in this sample to determine their multi-wavelength properties. We found that their IR and X-ray properties are consistent with their interpretation as powerful (bolometric luminosity $> 10^{45}$ erg sec$^{-1}$) obscured (hydrogen column densities $> 10^{22}$ cm$^{-2}$) AGN \citep{zaka04, ptak06}. 

Even though the direct emission from the nucleus is completely blocked by the obscuring material, some light emitted along other directions is scattered into our line of sight by electrons or dust particles in the host galaxy \citep{anto85}. The scattering efficiency (the ratio of the apparent luminosity due to scattering to the intrinsic luminosity) is very poorly known, but some observations suggest that it can be as large as a few per cent \citep{zaka05}. For an intrinsically luminous quasar such scattering efficiency implies that the scattered light can make a significant contribution to the observed flux. This contribution is very difficult to separate from the stellar light, since the scattered light can form complicated shapes in the plane of the sky. Imaging polarimetry helps resolve this ambiguity, but in the absence of such observations, spectropolarimetry or broad-band polarimetry allows one to determine the major direction of scattering and therefore indicates where in the galaxy most of the scattered light is coming from.

In this contribution we describe results of the imaging of nine type II quasars using the Advanced Camera for Surveys (ACS) aboard the Hubble Space Telescope (HST), in combination with polarimetric data from ground-based observations. We selected nine type II AGN with $L$([OIII]5007)$>10^9L_{\odot}$ from the samples by \citet{zaka03} and \citet{hao05} in the redshift range $0.25<z<0.4$. Based on the correlation between the narrow line luminosities and the broad-band luminosities that exists for unobscured quasars \citep{zaka03}, we estimate that the intrinsic luminosities of the objects in our program are about $-24>M_B>-26$. To avoid contamination from the strongest emission lines ([OII]3727, [OIII]4959,5007, H$\beta$, H$\alpha$), we carefully chose redshifts and ACS filters so that none of the filters used in our program contain any of these lines. 

\begin{figure}
\begin{center}
\includegraphics*[scale=0.77]{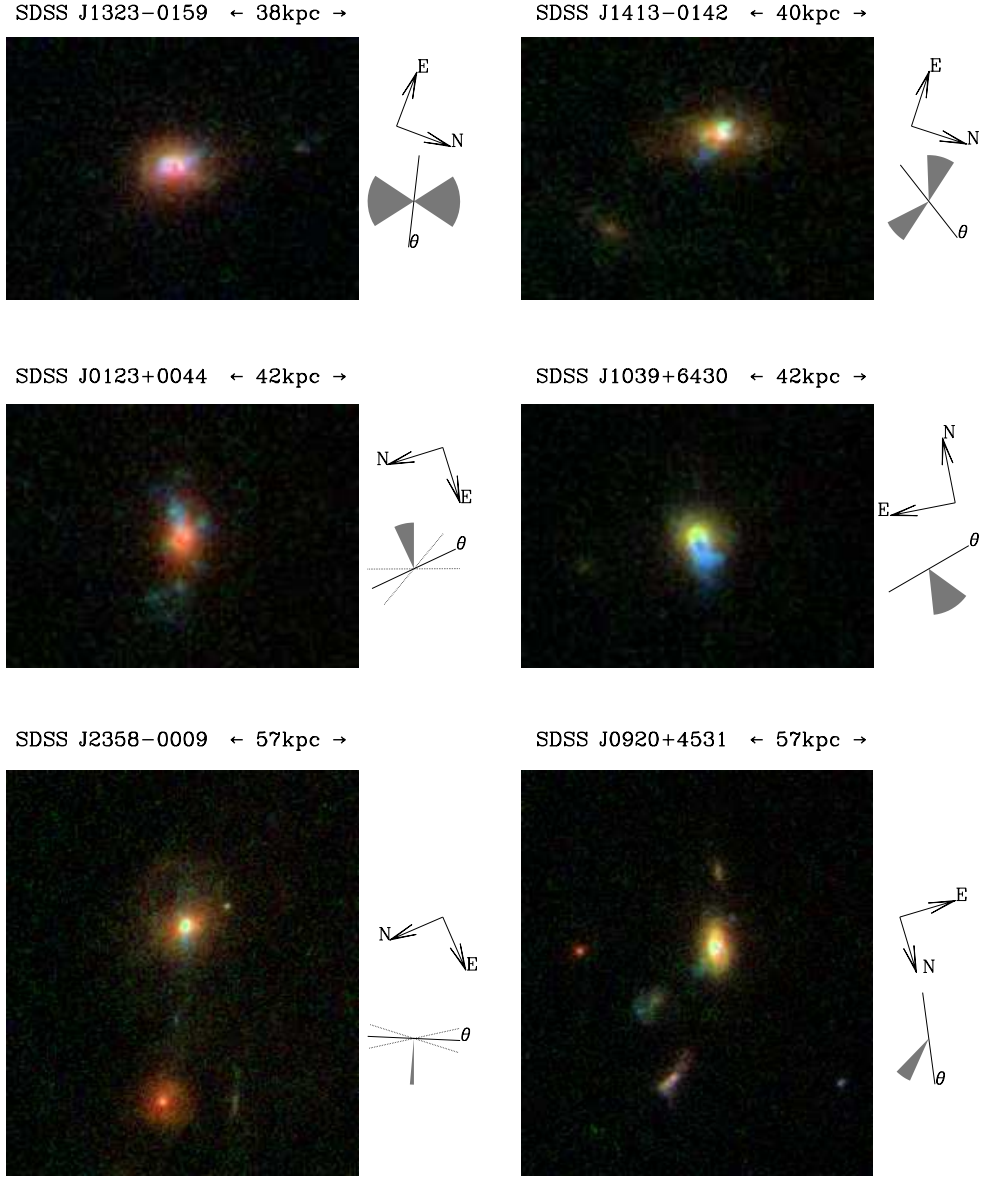}
\end{center}
\caption{Color-composite HST images of type II quasars. In all images similar parameters of the color-combining code were used, so the colors of the objects can be directly visually compared. Object identification and the total horizontal size of the images are given above each image. A cartoon of the scattering regions and polarization position angle (in image coordinates), as well as the orientation on the sky are shown to the right of each image. The polarization position angle $\theta$ is marked with a solid black line, and grey lines mark the $1\sigma$ confidence limits if the uncertainty in the angle is significant.}
\label{fig_1}
\end{figure}

\begin{figure}
\begin{center}
\includegraphics*[scale=0.7]{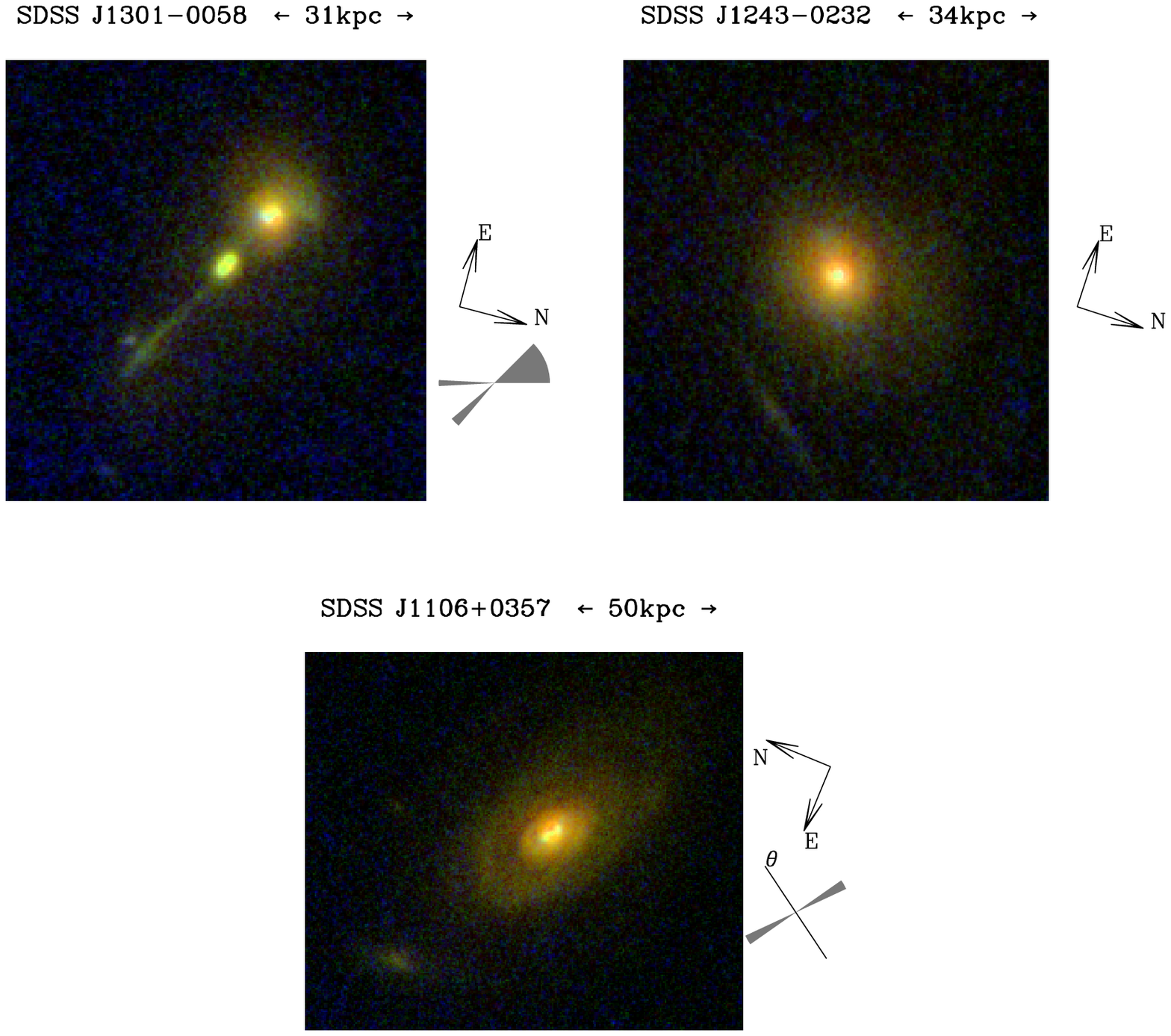}
\end{center}
\caption{Same as Figure \ref{fig_1}, except a different set of filters was used for these three objects. Polarimetric data are not available for SDSS~J1301$-$0058 and SDSS~J1243$-$0232.}
\label{fig_2}
\end{figure}

Each object was imaged in three filters: `blue' (blueward of the [OII]3727 line), `green' (between [OII]3727 and H$\beta$) and `red' (redward of the [OIII]5007 emission line). The three images were combined to produce color-composite images shown in Figures \ref{fig_1} and \ref{fig_2}. Within each Figure, the filters, redshifts and image combining parameters were similar, so that the difference in colors of objects in the images reflect differences in their optical spectral energy distributions. 

Polarimetric or spectropolarimetric data were obtained for seven of the objects using the CCD Spectropolarimeter (SPOL) at the 6.5m Multiple Mirror Telescope (MMT) and Bok telescope. The observed degrees of polarization vary from 1\% to 10\% (in the green band), and the polarization position angles are indicated with solid lines (marked `$\theta$') in Figures \ref{fig_1} and \ref{fig_2}. 

By fitting ellipses to the isophotes for each object in each band, we constructed one-dimensional brightness profiles and found that in most cases they are consistent with the de Vaucouleurs law, except SDSS~J1106+0357 (Figure \ref{fig_2}) which has a prominent exponential component in the outer parts. Once the best-fitting two-dimensional de Vaucouleurs profile is subtracted off the image, significant deviations are revealed. They are especially prominent in the blue band, and the brightest of them can be seen in Figures \ref{fig_1}-\ref{fig_2} as blue patches, either biconical (as in SDSS~J1323-0159) or irregularly shaped (as in SDSS~J1039+6430). We find that in most cases the line connecting the center of the galaxy and the brightest blue patch is orthogonal to the measured polarization position angle, and we therefore interpret these regions as emission from the obscured nucleus scattered into our line of sight by the material in the host galaxy.

Notable exceptions include SDSS~J0920+4531, in which the blue region near the center is misaligned with the scattering orientation as measured using spectropolarimetry. Given the overall morphology of the object which shows several components interacting with the main galaxy, it is likely that the bright blue patch near the center is just a starforming region rather than scattered light. In this case, the scattering region cannot be easily identified on the basis of morphology. The example of SDSS~J0920+4531 shows that in the absence of polarimetric data, star formation in the host cannot be easily distinguished from the scattered light contribution. This difficulty is further underscored by the case of SDSS~J0123+0044, where the ambiguity is not resolved even by the polarimetric data. In this object, several patches of blue surround the nucleus. While the brightest of them is orthogonal to the polarization position angle (and therefore is likely to be scattered rather than stellar light), the fainter spots to the east of the nucleus lie on the extension of a faint thin structure which extends out to 30 kpc to the north of the galaxy (this structure is too faint to be seen on Figure \ref{fig_1}). Given this morphological connection with what looks like a tidal tail, the eastern spots are probably patches of star formation in a companion in the process of disruption. 

Scattered light can make a significant contribution to the broad-band fluxes of quasar host galaxies. One striking example is SDSS~J1039+6430, in which most of the blue emission is due to the scattered light, and the measured polarization reaches 17\% in this band. Scattering from different parts of the host galaxy can wash out the polarization signal, so that scattered light can be important even if the observed polarization is not very high. If one were to take the colors of the extended emission at face value to estimate the age of the stellar population in the host galaxy, the measured ages would be strongly biased toward small values. As we demonstrated above, polarimetric data can help distinguish between scattered light and star formation, but does not eliminate the ambiguity entirely. This difficulty may be even more severe in host studies of unobscured quasars, in which the presence of a very bright blue nuclear source can make a detailed morphological analysis difficult and can dilute the polarization signal.

Our other findings include the presence of kpc-scale dust lanes across the centers of some host galaxies (e.g., SDSS~J1323$-$0159 and SDSS~J0123+0044), extended tidal structures in three of the objects in our sample, and faint companions within 20 projected kpc in each of the remaining six objects. The statistical significance of these results will be addressed in our future work. 

\section*{Acknowledgments}

NLZ is supported by a {\it Spitzer} fellowship. The observations reported here were obtained with the NASA/ESA Hubble Space Telescope and at the MMT Observatory, a facility operated jointly by the Smithsonian Institution and the University of Arizona. Public Access time is available at the MMT Observatory through an agreement with the National Science Foundation. Funding for the creation and distribution of the SDSS Archive has been provided by the Alfred P. Sloan Foundation, the Participating Institutions, NASA, the NSF, the U.S. Department of Energy, the Japanese Monbukagakusho, and the Max Planck Society. The SDSS Web site is http://www.sdss.org/. The SDSS is managed by the Astrophysical Research Consortium for the Participating Institutions. The Participating Institutions are The University of Chicago, Fermilab, the Institute for Advanced Study, the Japan Participation Group, The Johns Hopkins University, the Korean Scientist Group, Los Alamos National Laboratory, the Max-Planck-Institute for Astronomy (MPIA), the Max-Planck-Institute for Astrophysics (MPA), New Mexico State University, University of Pittsburgh, University of Portsmouth, Princeton University, the United States Naval Observatory, and the University of Washington.

\end{document}